\documentclass[10pt,twocolumn,letterpaper]{article}

\usepackage[pagenumbers]{cvpr} 

\usepackage{graphicx}
\usepackage{amsmath}
\usepackage{amssymb}
\usepackage{booktabs}
\usepackage{multirow}
\usepackage{subcaption}
\usepackage[colorlinks,linkcolor=red,citecolor=blue]{hyperref}

\usepackage[capitalize]{cleveref}
\crefname{section}{Sec.}{Secs.}
\Crefname{section}{Section}{Sections}
\Crefname{table}{Table}{Tables}
\crefname{table}{Tab.}{Tabs.}

\def\cvprPaperID{6890} 
\def\confName{CVPR}
\def\confYear{2022}

\begin{document}

\title{Semantic-guided Disentangled Representation for Unsupervised Cross-modality Medical Image Segmentation}

\author{Shuai Wang, Rui Li\\
Tsinghua University\\
{\tt\small \{s-wang20@mails.,leerui@\}tsinghua.edu.cn}
}
\maketitle

\begin{abstract}
   Disentangled representation is a powerful technique to tackle domain shift problem in medical image analysis in unsupervised domain adaptation setting.
   However, previous methods only focus on exacting domain-invariant feature and ignore whether exacted feature is meaningful for downstream tasks.
   We propose a novel framework, called semantic-guided disentangled representation (SGDR), an effective method to exact semantically meaningful feature for segmentation task to improve performance of cross modality medical image segmentation in unsupervised domain adaptation setting.
   To exact the meaningful domain-invariant features of different modality, we introduce a content discriminator to force the content representation to be embedded to the same space and a feature discriminator to exact the meaningful representation.
   We also use pixel-level annotations to guide the encoder to learn features that are meaningful for segmentation task.
   We validated our method on two public datasets and experiment results show that our approach outperforms the state of the art methods on two evaluation metrics by a significant margin.
\end{abstract}

\section{Introduction}
\label{sec:intro}

\begin{figure}[t]
\centering
\includegraphics[width=0.4\linewidth]{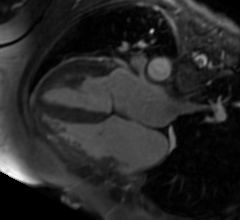}
\includegraphics[width=0.4\linewidth]{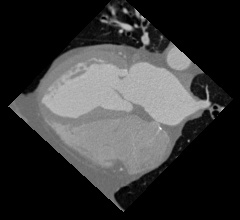}
\includegraphics[width=0.8\linewidth]{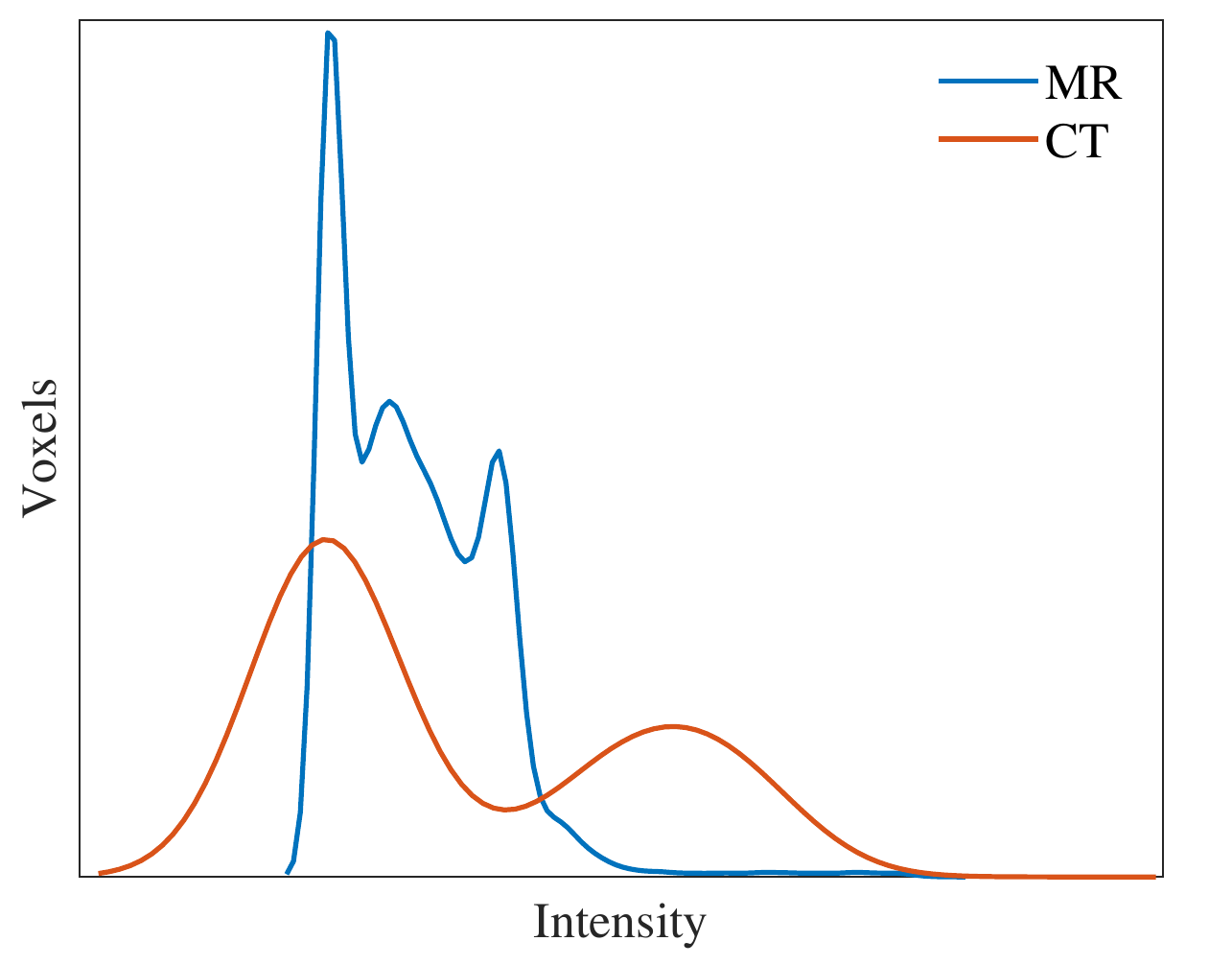}

\caption{Illustration of domain shift.Examples of MR and CT images from different subjects are shown in the first row. The second is intensity distribution of examples. In addition to the difference in gray distribution between MR and CT images, there is also an offset of anatomical structure}
\label{domainshift}
\end{figure}

Recently, deep learning methods have achieved great success in medical image analysis\cite{esteva2019guide}, such as segmentation \cite {unet,nnu,li2018h}, detection \cite {ZhengLGNC15,ueda2019deep,shi2020clinically} and so on.
There is a common assumption that training and test images are sampled from the sample distribution identically, \ie independently identically distribution.
However, in real-world scene, especially in medical image analysis, due to different imaging principles and acquisition parameters, a large domain gap between training data and test data usually occurs, as shown in \cref{domainshift}.
The large domain gap reduces the performance on test set and limits the application of deep learning model.

To reduce the domain gap and improve the performance, fine-tuning the model pre-trained on source domain with labeled data is a trivial way.
But it requires pixel-level annotations on the target domain, which is time-consuming and expensive.
Unsupervised domain adaptation (UDA) has attracted the attention in medical image analysis community \cite {kamnitsas2017unsupervised,sifa,Dou18, chiou2020harnessing,PeiWHZ21, ning2021new, li2020towards}.
In UDA setting, given labeled data from source domain and unlabeled data from target domain, our goal is to learn a model can segment the target domain in an unsupervised training manner.

For the existing UDA methods, there are two categories: distance-based methods \cite {MRCTdata2, tzeng2014deep, long2015learning, chanti2021olva, varda} and adversarial training methods \cite {ning2021new, PeiWHZ21, YangD0CLD19, JiangV20, chen2019unsupervised, kamnitsas2017unsupervised}.
Disentangled representation is often used in adversarial training methods, which is a powerful technique to exact the domain-invariant features and domain-specific features between source domain and target domain.
However, previous methods only focus on exacting features and do not consider whether exacted features are semantically meaningful for special task, such as segmentation, which leads to sub-optimal results.

By contrast, we propose a novel method to exact more meaningful features for our segmentation task.
First, we disentangle images of source domain and target domain onto content space and style space.
Content space includes domain-invariant information, such as anatomical structure and style space includes domain specific information, such as modality or contrast.
Generator learns to perform image-to-image translation combing the content features and style information.
Specially, we use a content discriminator to enforce the encoder exact the domain-invariant features between source domain and target domain.
To learn more meaningful and semantic-related features, we introduce a feature discriminator and auxiliary segmentation loss. 
We use cross cycle consistency loss to help our model can handle unpaired data.
Our model can be trained end to end with adversarial training.

Our contribution could be summarized as follows.
First, we present a novel unsupervised domain adaptation framework via disentangled representation.
Second, we introduce semantic-guided strategy to make exacted features more suitable for downstream tasks, such as semantic segmentation. Third, we conduct extensive experiments in the three cardiac segmentation tasks and experiment results show that our method consistently improves the performance in unsupervised domain adaptation setting and outperforms the state of the art approaches.
\section{Related Work}
\label{sec:formatting}

This work is mainly related to two areas: domain adaptation and disentangled representation.

\subsection{Domain Adaptation}

Domain adaptation is a potential technology for tackling domain shift problem in medical image analysis. Previous domain adaptation methods can be divided into two categorizes: a)minimizing the distance metrics to align the data in a common space explicitly. b) adversarial training which reduce the domain gap implicitly based on generative adversarial network \cite{gan}.

Tzeng \etal \cite {tzeng2014deep} and Long \etal \cite {long2015learning} minimized the Maximum Mean Discrepancy (MMD) to reduce the domain discrepancy. Sun \cite{sun2016deep} proposed COARL, which aligns the second-order statistics of the source domain and target domain. Courty \etal \cite{CourtyFHR17} proposed a novel method that minimizes the optimal transport loss between the joint source distribution and an estimated target joint distribution.
Kang \etal \cite{kang2020pixel} proposed pixel-level cycle association to bulid the connection between source and target pixel pairs, which is a new perspective for semantic segmentation under unsupervised domain adapation setting.
For medical image segmentation, Wu and Zhang \cite {MRCTdata2} proposed Characteristic Function Distance (CFD) to reduce the domain gap in frequency domain through Fourier Transformation. Chanti \cite{chanti2021olva} proposed a method that learns a shared cross-domain latent space by variational auto-encoder and optimal transport theory.

Generative adversarial network \cite{gan} is widely used in image generation and domain adaptation. The main idea is to align the source data and target data on image level or feature level by adversarial training. In unsupervised domain adaptation setting, some methods aim to learn domain-invariant features across domains \cite{dann,adaa, adaptsegnet, kamnitsas2017unsupervised, Dou18}, while other methods aligns the source domain image and target domain image on image level \cite{chiou2020harnessing, chen2019unsupervised, JiangV20} based some image to image networks, such as CycleGAN \cite{cyclegan}, MUNIT \cite{MUNIT} or  DRIT \cite{DRIT}. Chen \etal \cite{sifa} presented a novel framework aligns source domain and target domain on both image level and feature level, which achieves the state of the art performance.

\subsection{Disentangled Representation}

Disentangled representation aims to learn domain-invariant feature and domain-specific feature simultaneously from paired \cite{Gonzalez-Garcia18} or unpaired data \cite{MUNIT,DRIT}. InfoGAN \cite{infogan} learns disentanglement representation by maximizing the variational upper bound of mutual information between latent variables and data variation. Disentangling multimodal images has been used for liver segmentation with domain adaptation \cite{YangD0CLD19}.
Bousmails \etal \cite{bousmalis2016domain} explicitly separated domain-invariant features and domain-specific by encouraging orthogonality between shared and specific of each domain.
Pei \etal \cite{PeiWHZ21} introduce zero-loss \cite{zeroloss} to enhance the disentanglement for cross-modality cardiac segmentation. However, previous methods only focus on extracting domain-invariant features, and ignore whether the extracted features are meaningful for downstream tasks, such as classification, detection and segmentation. This leads to performance degradation. To tackle this problem, we propose semantic-guided disentangled representation by using pixel-level annotations of source domain and two discriminators to help capturing the meaningful representation.


\begin{figure*}[t]
\centering
\includegraphics[width=0.8\linewidth]{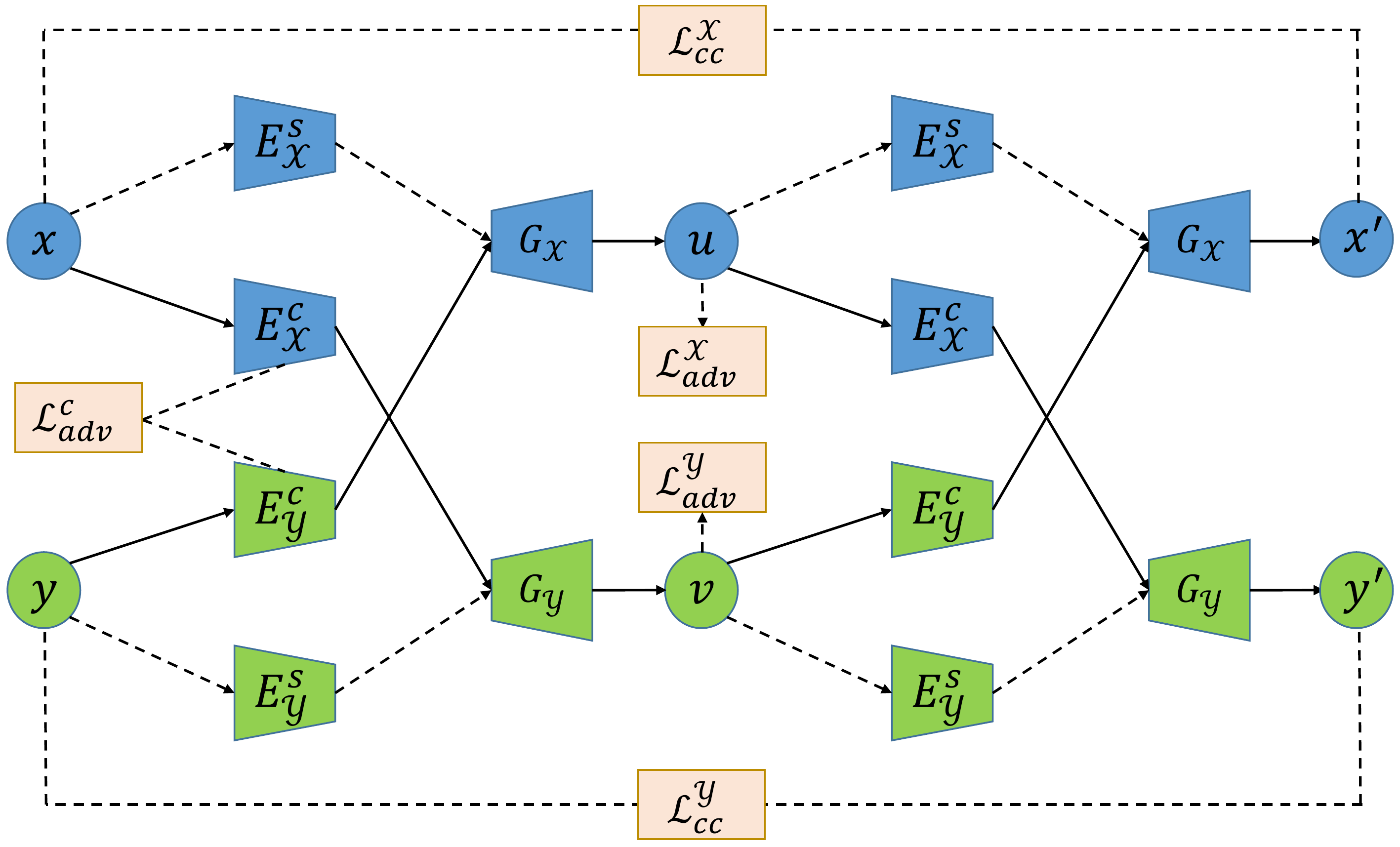}
\caption{Our framework overview (\cref{sec:semantic}).
$x$ and $y$ are sampled images from source domain $\mathcal{X}$ and target domain $\mathcal{Y}$ respectively.
$\{E^c_{\mathcal{X}},E^c_{\mathcal{Y}}\}$ and $\{E^s_{\mathcal{X}},E^s_{\mathcal{Y}}\}$ are corresponding content encoders and corresponding sytle encoders.
$\{G_{\mathcal{X}},G_{\mathcal{Y}}\}$ are image generators.
$u$ and $v$ are translated images of source domain and target domain.
$x'$ and $y'$ are cross cycle reconstructed images of $x$ and $y$.}
\label{network1}
\end{figure*}

\section{Method}
An overview of our method is shown in \cref{network1}. 
We first disentangle image of each domain onto two space: domain-invariant space $\mathcal{C}$ and domain specific space $\mathcal{S_X}$ and $\mathcal{S_Y}$ using four encoders.
After that, we swap the domain-invariant features of each domain and feed them into generators to get fake images.
And we repeat the encoding and generation on the generated fake images.
Moreover, we use a content discriminator $D^c$ to get the domain-invariant features.
To make exacted features more meaningful and suitable for segmentation tasks, we use semantic-guided representation learning strategy.
More details will be discussed in \cref{sec:semantic}.
The other parts are organized as follows.
We introduce our problem setting, \ie unsupervised domain adaptation for medical image segmentation in \cref{definition}.
The total objective function and implement details could be found in \cref{loss} and \cref{implement}.
\subsection{Problem Definition}
\label{definition}
We consider semantic segmentation of unsupervised domain adaptation. Let denote  labeled data from source domain as $\{(x^i,x^i_{gt})|i=1,2,3 \cdots,M\} $ and unlabeled data from target domain as $\{y^i|i=1,2,3 \cdots,N\}$, respectively, where $x^i$ and $y^i$ denote images from source domain and target domain and $x^i_{gt}$ is pixel-level annotation of $x^i$. Our goal is to train a model from labeled data from source domain data and unlabeled data from target domain and the model can segment data from target domain directly.
\subsection{ Semantic-guided Disentangled Representation}
\label{sec:semantic}
Let $x\in\mathcal{X}$ and $y\in\mathcal{Y}$ denote unpaired images are sampled from source domain $\mathcal{X}$ and target domain $\mathcal{Y}$.
Our method embeds images to content space $\mathcal{C}$ and style space: $\mathcal{S_X}$ and $\mathcal{S_Y}$. Content space includes domain-invariant features and style space includes domain specific features.
Intuitively, in terms of medical images, the content space include domain-invariant features such as anatomical structure and the style space includes modal information such as contrast.
The content encoder $E^c_{\mathcal{X}}$ or $E^c_{\mathcal{Y}}$ should exact the shared information between the source domain and target domain, while style encoder $E^s_{\mathcal{X}}$ or $E^s_{\mathcal{Y}}$ encodes the domain specific information onto $\mathcal{S_X}$ and $\mathcal{S_Y}$.

To facilitate the factorization of content and style feature, we use a content discriminator $D^c$ to make the encoders of source domain and target domain, \ie $E^c_{\mathcal{X}}$ and $E^c_{\mathcal{Y}}$ , exact the domain-invariant features.
The content discriminator $D^c$ aims to distinguish the content representation from source domain or target domain. The content encoders learn to exact domain invariant features that can’t be distinguished by the content discriminator. We formulate this adversarial training process as follows:
\begin{equation}
\begin{split}
    \mathcal{L}^c_{adv} = \mathbb{E}_x[\log D^c(E^c_{\mathcal{X}}(x))+\log(1-D^c(E^c_{\mathcal{X}}(x)))] \\
                + \mathbb{E}_y[\log D^c(E^c_{\mathcal{Y}}(y))+\log(1-D^c(E^c_{\mathcal{Y}}(y)))] 
\end{split}
\end{equation}
\begin{figure}[t]
\centering
\includegraphics[width=0.8\linewidth]{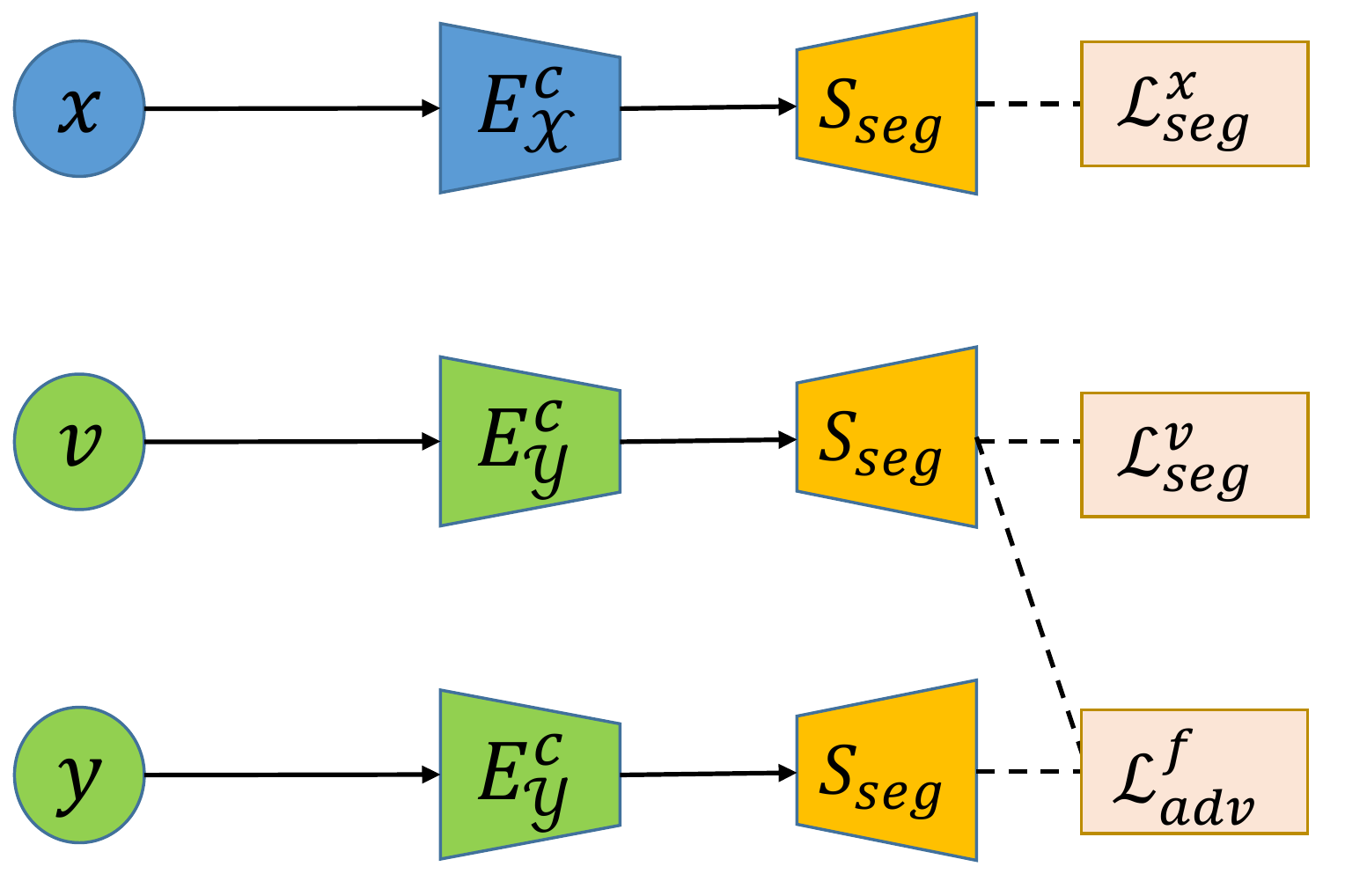}
\caption{Illustration of our semantic-guided strategy.
$S_{seg}$ is segmentation module to get segmentation prediction.
In order to capture meaningful representation from input images, we add segmentation loss to content encoder $E^c_{\mathcal{X}}$ and segmentation module $S_{seg}$.
As anatomical structure is similar between the source domain and target domain, we introduce a feature discriminator to force the anatomical structure of target domain to be as close to source domain.}
\label{network2}
\end{figure}
However, domain-invariant information is not always meaningful for semantic segmentation. Moreover, meaningless information will cause performance degradation.
We introduce a feature discriminator and an auxiliary loss function to help the encoder get more meaningful information for segmentation. First, we add a segmentation network $S_{seg}$ after content encoder, as shown in \cref{network2}. We can train content encoder $E^c_{\mathcal{X}}$ and segmentation network $S_{seg}$ together with source domain label to make content features more meaningful for segmentation task. The segmentation loss is as follows:
\begin{equation}
\mathcal{L}^x_{seg} = Dice(\widetilde{x},x_{gt}) + CE(\widetilde{x},x_{gt})
\end{equation}
where $\widetilde{x}=S_{seg}(E^c_{\mathcal{X}}(x))$ is segmentation prediction of source domain.
$Dice$ represents Dice loss and the second term $CE$ is cross-entropy loss.
We also add an auxiliary segmentation loss $L^{v}_{seg}$ to constrain the anatomical shape of fake images as shown in \cref{network2}.
The segmentation loss is defined as:
\begin{equation}
\begin{split}
\mathcal{L}^v_{seg} = Dice(\widetilde{v},x_{gt}) + CE(\widetilde{v},x_{gt})
\end{split}
\end{equation}
where $\widetilde{v}=S_{seg}(E^c_{\mathcal{Y}}(v))$. The total segmentation loss is formulated as follows:
\begin{equation}
\mathcal{L}_{seg} = \mathcal{L}^x_{seg} + \mathcal{L}^v_{seg}
\end{equation}

In order to improve performance on target domain and exact more meaningful features, we introduce feature adversarial training for content encoders $E^c_{\mathcal{Y}}$ and segmentation module $S_{seg}$, which could be view as semantic-level supervision.
It is noted that anatomical structures are similar between source domain and target domain, so the feature discriminator would fail to distinguish segmentation masks from source domain or target domain.
The adversarial training loss could be formulated as follows:

\begin{equation}
\begin{split}
\mathcal{L}^f_{adv}=\mathbb{E}_v[\log D^f(S_{seg}(E^c_{\mathcal{Y}}(v))))] + \\
                    \mathbb{E}_y[\log (1-D^f(S_{seg}(E^c_{\mathcal{Y}}(y)))))]
\end{split}
\end{equation}
where $D^f$ is a discriminator to distinguish segmentation mask comes from $y$ or $v$.

For style encoding, we follow the common approach in \cite {chartsias2020disentangle, DRIT, chen2019robust} and we set style encoding $s\in R^8$ assuming its posterior distribution $q(s|x)$ to be a Gaussian distribution $\mathcal{N}(0,I)$. We use Kullback-Leibler (KL) divergence to force $q(s|x)$ to be close to the normal distribution for training style encoder, the KL loss is formulated as follows:
\begin{equation}
\mathcal{L}_{KL}=D_{KL}(q(s|x)||\mathcal{N}(0,I))
\end{equation}
where $D_{KL}(p||q)=\int p(x)\ln(\frac{p(x)}{q(x)}){\rm d}x$.

After disentangling images to content space and style space, we combine the content feature and style feature into generator $G_{\mathcal{X}}$ or $G_{\mathcal{Y}}$ and get a fake image $u=G_{\mathcal{X}}(E^c_{\mathcal{Y}}(y),E^s_{\mathcal{X}}(x))$ or $v=G_{\mathcal{Y}}(E^c_{\mathcal{X}}(x),E^s_{\mathcal{Y}}(y))$.
$u$ has the same anatomical structure with $y$ and same style with $x$, \ie $u$ and $x$ should come from the same domain. We use adversarial training to generate more realistic image of source domain or target domain. We denote adversarial training loss as $\mathcal{L}^{\mathcal{X}}_{adv}$ and $\mathcal{L}^{\mathcal{Y}}_{adv}$.
For $\mathcal{L}^{\mathcal{X}}_{adv}$ and $\mathcal{L}^{\mathcal{Y}}_{adv}$, we use a least-square loss\cite{mao2017least}, which is more stable during training process and generates more realistic images.
In particular, for $\mathcal{L}^{\mathcal{X}}_{adv}$, we train the $G_{\mathcal{X}}$ to minimize $\mathbb{E}_{u}[(D_{\mathcal{X}}(u)-1)^2]$ and train the $D_{\mathcal{X}}$ to minimize $\mathbb{E}_{u}[D_{\mathcal{X}}(u)^2] +\mathbb{E}_{x}[(D_{\mathcal{X}}(x)-1)^2]$.
We can also formulate $\mathcal{L}^{\mathcal{Y}}_{adv}$ symmetrically.
For convenience, we define:
\begin{equation}
\mathcal{L}^{domain}_{adv}=\mathcal{L}^{\mathcal{X}}_{adv} + \mathcal{L}^{\mathcal{Y}}_{adv}
\end{equation}

We repeat the encoding and generation operations again on the generated images.
Finally, we get $x'=G_{\mathcal{X}}(E^c_{\mathcal{Y}}(v),E^s_{\mathcal{X}}(u))$ and $y'=G_{\mathcal{Y}}(E^c_{\mathcal{X}}(u),E^s_{\mathcal{Y}}(v))$, which should be identical to the original ones, such as $x$ or $y$.
Inspired by \cite {cyclegan, DRIT}, we use cross-cycle-consistency loss to enforce this constraint. This loss could be formulated as:
\begin{equation}
\mathcal{L}_{cc}=\mathbb{E}_x \Vert x'-x \Vert_1 + \mathbb{E}_y \Vert y'-y \Vert_1
\end{equation}

\subsection{Objective Function}
\label{loss}
We also use other loss functions to improve our framework.

\textbf{Self-reconstruction loss}. When $x$ is encoded to content feature and style feature, the generator $G_{\mathcal{X}}$ should recover $x$.
Let denote $\hat{x}=G_{\mathcal{X}}(E^c_{\mathcal{X}}(x),E^s_{\mathcal{X}}(x))$ and $\hat{x}$ should be identical to $x$, we formulate this constrain as $\mathcal{L}^{\mathcal{X}}_{recon}=\mathbb{E}_x\Vert \hat{x}-x \Vert_1$. We also formulate $\mathcal{L}^{\mathcal{Y}}_{recon}$ symmetrically. Self-reconstruction loss ensures that we have exacted all information we need. The self-reconstruction loss could be formulated as follows totally:
\begin{equation}
\mathcal{L}_{recon}=\mathcal{L}^{\mathcal{X}}_{recon}+\mathcal{L}^{\mathcal{Y}}_{recon}
\end{equation}

\textbf{Latent regression loss}. Similar to BicycleGAN \cite{bicyclegan}, we add latent regression loss to encourage invertible mapping between the real images and fake images. Latent regression loss also avoids posterior collapse of style encode and prevent generator from ignoring style information \cite {chartsias2020disentangle}. We minimize the following loss:
\begin{equation}
\begin{split}
\mathcal{L}_{latent}=\mathbb{E}_x\Vert z- E^s_{\mathcal{X}}(G_{\mathcal{X}}(z,E^c_{\mathcal{X}}(x)))\Vert_1 + \\
\mathbb{E}_y\Vert z- E^s_{\mathcal{Y}}(G_{\mathcal{Y}}(z,E^c_{\mathcal{Y}}(y)))\Vert_1
\end{split}
\end{equation}
where $z$ is sampled from gaussian distribution $\mathcal{N}(0,I)$.

\begin{figure*}[t]
\centering
MR to CT \\
\begin{subfigure}{0.13\linewidth}
\includegraphics[scale=0.3]{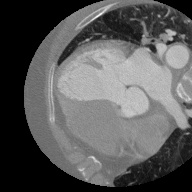}
\end{subfigure}
\begin{subfigure}{0.13\linewidth}
\includegraphics[scale=0.3]{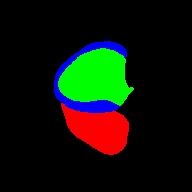}
\end{subfigure}
\begin{subfigure}{0.13\linewidth}
\includegraphics[scale=0.3]{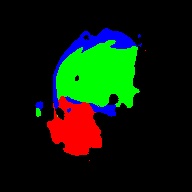}
\end{subfigure}
\begin{subfigure}{0.13\linewidth}
\includegraphics[scale=0.3]{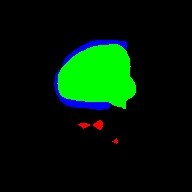}
\end{subfigure}
\begin{subfigure}{0.13\linewidth}
\includegraphics[scale=0.3]{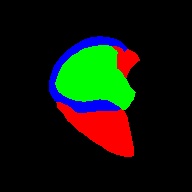}
\end{subfigure}
\begin{subfigure}{0.13\linewidth}
\includegraphics[scale=0.3]{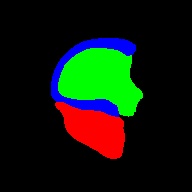}
\end{subfigure}
\begin{subfigure}{0.13\linewidth}
\includegraphics[scale=0.3]{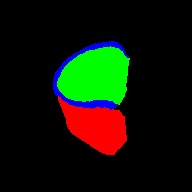}
\end{subfigure}

\begin{subfigure}{0.13\linewidth}
\includegraphics[scale=0.3]{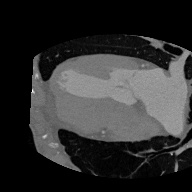}
\end{subfigure}
\begin{subfigure}{0.13\linewidth}
\includegraphics[scale=0.3]{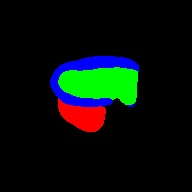}
\end{subfigure}
\begin{subfigure}{0.13\linewidth}
\includegraphics[scale=0.3]{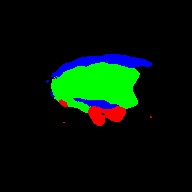}
\end{subfigure}
\begin{subfigure}{0.13\linewidth}
\includegraphics[scale=0.3]{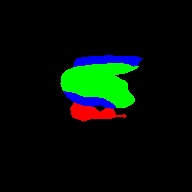}
\end{subfigure}
\begin{subfigure}{0.13\linewidth}
\includegraphics[scale=0.3]{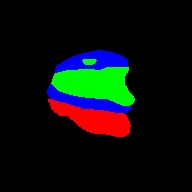}
\end{subfigure}
\begin{subfigure}{0.13\linewidth}
\includegraphics[scale=0.3]{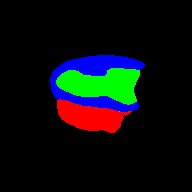}
\end{subfigure}
\begin{subfigure}{0.13\linewidth}
\includegraphics[scale=0.3]{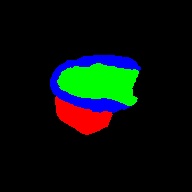}
\end{subfigure}

\begin{subfigure}{0.13\linewidth}
\includegraphics[scale=0.3]{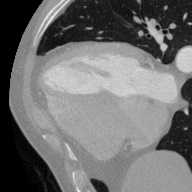}
\end{subfigure}
\begin{subfigure}{0.13\linewidth}
\includegraphics[scale=0.3]{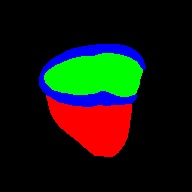}
\end{subfigure}
\begin{subfigure}{0.13\linewidth}
\includegraphics[scale=0.3]{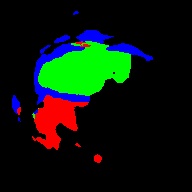}
\end{subfigure}
\begin{subfigure}{0.13\linewidth}
\includegraphics[scale=0.3]{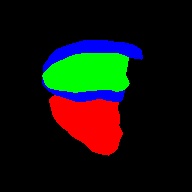}
\end{subfigure}
\begin{subfigure}{0.13\linewidth}
\includegraphics[scale=0.3]{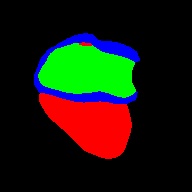}
\end{subfigure}
\begin{subfigure}{0.13\linewidth}
\includegraphics[scale=0.3]{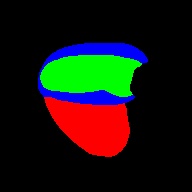}
\end{subfigure}
\begin{subfigure}{0.13\linewidth}
\includegraphics[scale=0.3]{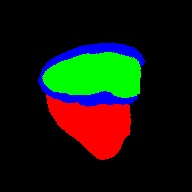}
\end{subfigure}

CT to MR \\

\begin{subfigure}{0.13\linewidth}
\includegraphics[scale=0.3]{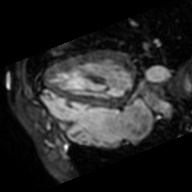}
\end{subfigure}
\begin{subfigure}{0.13\linewidth}
\includegraphics[scale=0.3]{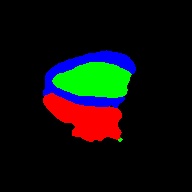}
\end{subfigure}
\begin{subfigure}{0.13\linewidth}
\includegraphics[scale=0.3]{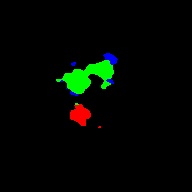}
\end{subfigure}
\begin{subfigure}{0.13\linewidth}
\includegraphics[scale=0.3]{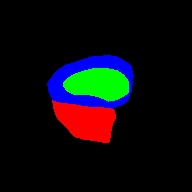}
\end{subfigure}
\begin{subfigure}{0.13\linewidth}
\includegraphics[scale=0.3]{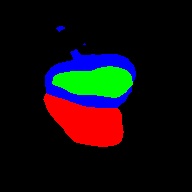}
\end{subfigure}
\begin{subfigure}{0.13\linewidth}
\includegraphics[scale=0.3]{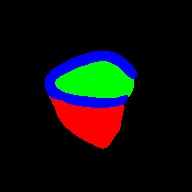}
\end{subfigure}
\begin{subfigure}{0.13\linewidth}
\includegraphics[scale=0.3]{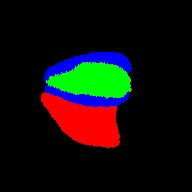}
\end{subfigure}

\begin{subfigure}{0.13\linewidth}
\includegraphics[scale=0.3]{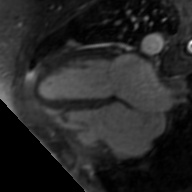}
\end{subfigure}
\begin{subfigure}{0.13\linewidth}
\includegraphics[scale=0.3]{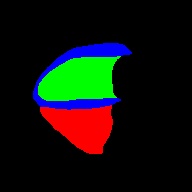}
\end{subfigure}
\begin{subfigure}{0.13\linewidth}
\includegraphics[scale=0.3]{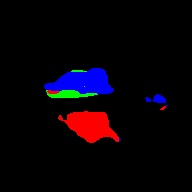}
\end{subfigure}
\begin{subfigure}{0.13\linewidth}
\includegraphics[scale=0.3]{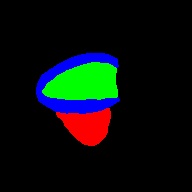}
\end{subfigure}
\begin{subfigure}{0.13\linewidth}
\includegraphics[scale=0.3]{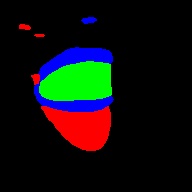}
\end{subfigure}
\begin{subfigure}{0.13\linewidth}
\includegraphics[scale=0.3]{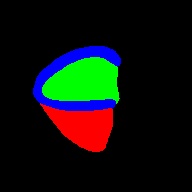}
\end{subfigure}
\begin{subfigure}{0.13\linewidth}
\includegraphics[scale=0.3]{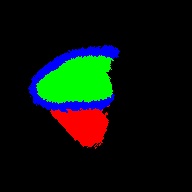}
\end{subfigure}

\begin{subfigure}{0.13\linewidth}
\includegraphics[scale=0.3]{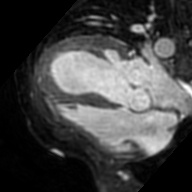}
\end{subfigure}
\begin{subfigure}{0.13\linewidth}
\includegraphics[scale=0.3]{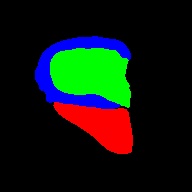}
\end{subfigure}
\begin{subfigure}{0.13\linewidth}
\includegraphics[scale=0.3]{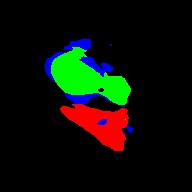}
\end{subfigure}
\begin{subfigure}{0.13\linewidth}
\includegraphics[scale=0.3]{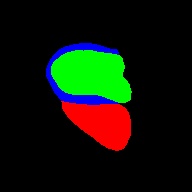}
\end{subfigure}
\begin{subfigure}{0.13\linewidth}
\includegraphics[scale=0.3]{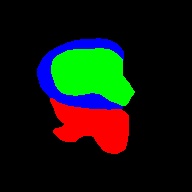}
\end{subfigure}
\begin{subfigure}{0.13\linewidth}
\includegraphics[scale=0.3]{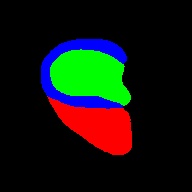}
\end{subfigure}
\begin{subfigure}{0.13\linewidth}
\includegraphics[scale=0.3]{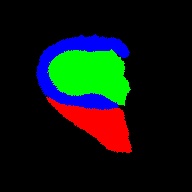}
\end{subfigure}

bSSFP to LGE\\

\begin{subfigure}{0.13\linewidth}
\includegraphics[scale=0.3]{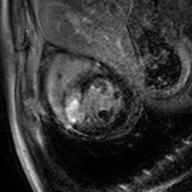}
\end{subfigure}
\begin{subfigure}{0.13\linewidth}
\includegraphics[scale=0.3]{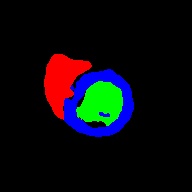}
\end{subfigure}
\begin{subfigure}{0.13\linewidth}
\includegraphics[scale=0.3]{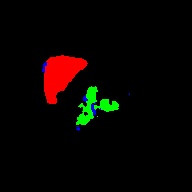}
\end{subfigure}
\begin{subfigure}{0.13\linewidth}
\includegraphics[scale=0.3]{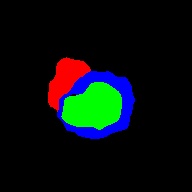}
\end{subfigure}
\begin{subfigure}{0.13\linewidth}
\includegraphics[scale=0.3]{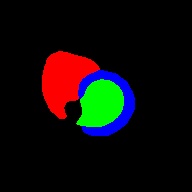}
\end{subfigure}
\begin{subfigure}{0.13\linewidth}
\includegraphics[scale=0.3]{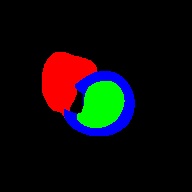}
\end{subfigure}
\begin{subfigure}{0.13\linewidth}
\includegraphics[scale=0.3]{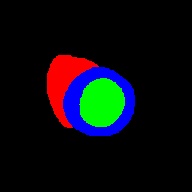}
\end{subfigure}

\begin{subfigure}{0.13\linewidth}
\includegraphics[scale=0.3]{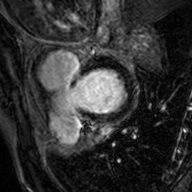}
\end{subfigure}
\begin{subfigure}{0.13\linewidth}
\includegraphics[scale=0.3]{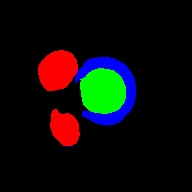}
\end{subfigure}
\begin{subfigure}{0.13\linewidth}
\includegraphics[scale=0.3]{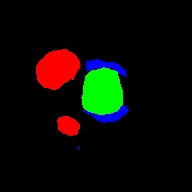}
\end{subfigure}
\begin{subfigure}{0.13\linewidth}
\includegraphics[scale=0.3]{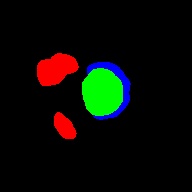}
\end{subfigure}
\begin{subfigure}{0.13\linewidth}
\includegraphics[scale=0.3]{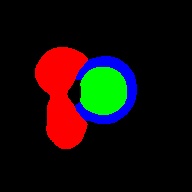}
\end{subfigure}
\begin{subfigure}{0.13\linewidth}
\includegraphics[scale=0.3]{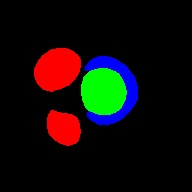}
\end{subfigure}
\begin{subfigure}{0.13\linewidth}
\includegraphics[scale=0.3]{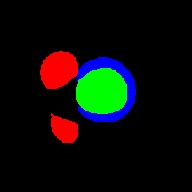}
\end{subfigure}

\begin{subfigure}{0.13\linewidth}
\includegraphics[scale=0.3]{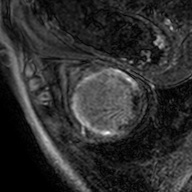}
\caption*{image}
\end{subfigure}
\begin{subfigure}{0.13\linewidth}
\includegraphics[scale=0.3]{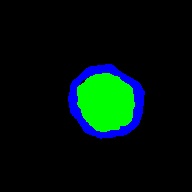}
\caption*{Unet-sup}
\end{subfigure}
\begin{subfigure}{0.13\linewidth}
\includegraphics[scale=0.3]{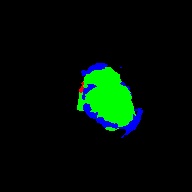}
\caption*{Unet-noada}
\end{subfigure}
\begin{subfigure}{0.13\linewidth}
\includegraphics[scale=0.3]{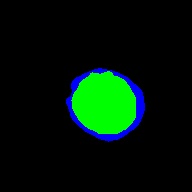}
\caption*{AdaptSeg}
\end{subfigure}
\begin{subfigure}{0.13\linewidth}
\includegraphics[scale=0.3]{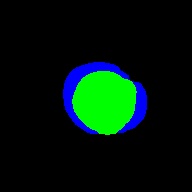}
\caption*{SIFA}
\end{subfigure}
\begin{subfigure}{0.13\linewidth}
\includegraphics[scale=0.3]{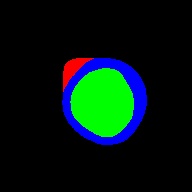}
\caption*{Ours}
\end{subfigure}
\begin{subfigure}{0.13\linewidth}
\includegraphics[scale=0.3]{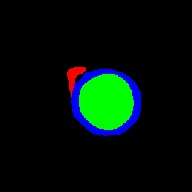}
\caption*{GT}
\end{subfigure}
\caption{Visual comparison of segmentation results of different methods.
The structures MYO, LV and RV are indicated in blue,green and red respectively.
From left to right: raw data sampled from target domain(1st column), results of different compared models (from 2nd column to 4th column), our results (5th column), ground truth of raw data (last column).}
\label{visual comparsion}
\end{figure*}

\textbf{Total objection function}. Totally, the full objective function of our method is as follows:
\begin{equation}
\begin{split}
\mathcal{L} = \lambda_{cc}\mathcal{L}_{cc} + \lambda_{recon}\mathcal{L}_{recon} + \lambda^c_{adv}\mathcal{L}^c_{adv} \\
+ \lambda^{domain}_{adv}\mathcal{L}^{domain}_{adv}+\lambda_{seg}\mathcal{L}_{seg} + \lambda^f_{adv}\mathcal{L}^f_{adv} \\
+ \lambda_{KL}\mathcal{L}_{KL} + \lambda_{latent}\mathcal{L}_{latent}
\end{split}
\end{equation}
where the hyper parameter $\lambda$ controls the importance of each term. In practice, we set $\lambda_{cc}=\lambda_{recon}=\lambda_{latent}=10$, $\lambda^c_{adv}=\lambda^{domain}_{adv}=\lambda_{seg}=\lambda^f_{adv}=1$ and $\lambda_{KL}=0.01$.

At test time, images from target domain are forwarded into content encoder $E^c_{\mathcal{Y}}$ and segmentation module $S_{seg}$.
In this way, we can obtain segmentation results $\widetilde{y}=S_{seg}(E^c_{\mathcal{Y}}(y))$.
\subsection{Implement details}
\label{implement}
In this section, we describe the network architecture of proposed method. We implemented our framework in PyTorch 1.8.1 on a RTX 2080Ti GPU. For content encoder, we use an architecture consisting of 6 convolution layers with an instance normalization layer and 8 residual blocks \cite {resnet}.
For style encoder, we use a CNN architecture with 5 convolution layers and full connected layers.
The style encode is set to $s\in R^8$.
For generator, we use 3 residual blocks and 3 nearest-neighbor upsampling layers to get the same resolution with input images. We share the weight between the last 3 layers of $E^c_{\mathcal{X}}$ and $E^c_{\mathcal{Y}}$ and the first layers of $G_{\mathcal{X}}$ and $G_{\mathcal{Y}}$.
For segmentation network $S_{seg}$, we use 4 residual blocks and 3 upsampling layers to get segmentation result. For all discriminators, we follow the configuration of PatchGAN proposed in \cite{patchgan}.
For training, we used the Adam optimizer with a batch size of 4 and with a learning rate of 0.001 for segmentation network and 0.0001 for other networks.
We keep the same learning rate for the first 150 epochs and linearly decay the learning rate to zero over the next 150 epochs.

\section{Experiments}
\subsection{Datasets and Preprocessing}
We evaluated our method on two public datasets. The first dataset is MR-CT dataset\footnote{\href{https://github.com/FupingWu90/CT\_MR\_2D\_Dataset\_DA}{https://github.com/FupingWu90/CT\_MR\_2D\_Dataset\_DA}}, an open dataset mainly from Multi-Modality Whole Heart Segmentation Challenge \cite {MRCTdata1, MRCTdata2}.
The dataset includes 52 CT images and 46 MR images from different subjects. Only 20 subjects of dataset have gold standard segmentation for MR or CT data. We use 10 subjects of them for test and  other images are used for training. We exact 16 slices for every subject from long-axis view around the heart to train our model.
The second dataset\footnote{\href{http://www.sdspeople.fudan.edu.cn/zhuangxiahai/0/mscmrseg19/}{http://www.sdspeople.fudan.edu.cn/zhuangxiahai/0/mscmrseg19/}} is from Multi-sequence Cardiac MR Segmentation (MSCMR) Challenge (MICCAI 2019)\cite{mscmr1,mscmr2}, which contains three sequences cardicac MR images from 45 subjects. 
We use two sequence images from each subjects: bSSFP and LGE images.
25 LGE and 45 bSSFP images are used for training and 20 LGE images are used for evaluation. Note that the LGE and bSSFP images are collected from the same patient, so we shuffle the dataset to make them unpaired.

For all datasets, we resample images to $1\times 1$ mm slice by slice and crop them to $192\times 192$. We convert images to $[-1,1]$ before feeding into the network. The data augmentations include random rotation and flipping. 
\subsection{Experiment Details and Evaluation Metrics}
We trained our model from scratch and evaluated our method on two datasets as mentioned.
For MR-CT dataset, we evaluated our method in two directions, \ie setting CT as source domain and MR as target domains and setting MR as source domain and CT as target domain in turn. For MSCMR dataset, we only evaluate on one direction, \ie bSSFP images as source domain and LGE images as target domain. We aim to segment the myocardium of left ventricle (MYO), left ventricular cavity (LV), and right ventricular cavity (RV) for two datasets.

For evaluation, we used two commonly-used metrics for medical image segmentation: Dice coefficient ([\%]) and average symmetric surface distance (ASSD). Dice coefficient measures overlap between prediction and ground truth and ASSD evaluates the segmentation performance at object boundary.
We evaluated our method slice by slice on both MR-CT dataset and MSMCR dataset.
Note that higher Dice coefficient and lower ASSD mean better performance.

\begin{table*}
  \centering
  \caption{Performance comparison between our model and other unsupervised domain adaption methods on three tasks. We evaluate performance from two metrics: Dice and ASSD for all cardiac structure and average.Note that higher Dice and lower ASSD mean better performance.}
  \begin{tabular}{lcccccccc}
  \toprule 
  \multicolumn{9}{l}{MR$\longrightarrow$ CT}\\
  \hline
  \multirow{2}*{Method} &  \multicolumn{4}{c}{Dice(\%)} & \multicolumn{4}{c}{ASSD(mm)} \\ 
  \cmidrule(lr){2-5}\cmidrule(lr){6-9} & MYO & LV & RV & Average & MYO & LV & RV & Average\\
  \hline
       Unet(supervised)\cite{unet} & 86.89 & 93.04 & 90.75 & 90.27 & 1.24 & 1.33 & 1.85 & 1.47 \\
       Unet(no adaptation)\cite{unet} & 29.88 & 74.87 & 51.99 & 52.24 & 6.54 & 5.96 &8.55 & 7.02 \\
       AdaptSeg \cite{adaptsegnet} & 47.65 & 77.41 & 64.43 & 63.16 & 3.78 & 3.41 & 5.18 & 4.12 \\
       SIFA \cite{sifa} & 52.94 & 82.32 & 83.62 & 72.96 & 3.19 & 3.88 & 4.00 & 3.69 \\
       Ours & 69.26 & 85.86 & 85.42 & 80.18 & 2.36 & 2.69 & 3.70 & 2.58 \\
  \hline
  \multicolumn{9}{l}{CT$\longrightarrow$ MR}\\
  \hline
  \multirow{2}*{Method} & \multicolumn{4}{c}{Dice(\%)} & \multicolumn{4}{c}{ASSD(mm)} \\ 
  \cmidrule(lr){2-5}\cmidrule(lr){6-9} & MYO & LV & RV & Average & MYO & LV & RV & Average\\
  \hline
     Unet(supervised)\cite{unet} & 82.84 & 93.86 & 90.81 & 89.17 & 1.46 & 1.69 &1.93 &1.69 \\
     Unet(no adaptation)\cite{unet} & 14.84 & 53.67 & 62.13 & 43.55 & 10.85 & 8.88 & 6.69 & 8.80 \\
     AdaptSeg \cite{adaptsegnet} & 47.80 & 74.30 &73.30 & 62.13 & 4.80 & 4.66 & 4.33 &4.59 \\
     SIFA \cite{sifa} & 65.30 & 78.60 & 76.00 & 73.30 & 5.20 & 4.60 & 6.20 & 5.30 \\
     Ours & 65.42 & 83.66 & 79.20 & 76.10 & 3.55 & 4.17 & 4.66 & 4.12 \\
  \hline
  \multicolumn{9}{l}{bSSFP$\longrightarrow$ LGE}\\
  \hline
  \multirow{2}{*}{Method} & \multicolumn{4}{c}{Dice(\%)} & \multicolumn{4}{c}{ASSD(mm)} \\ 
  \cmidrule(lr){2-5}\cmidrule(lr){6-9} & MYO & LV & RV & Average & MYO & LV & RV & Average\\
  \hline
    Unet(supervised)\cite{unet}& 77.46 & 86.49 & 70.91 & 78.29 & 1.67 & 1.52 & 2.37 &1.85 \\
    Unet(no adaptation)\cite{unet} & 33.77 & 62.03 &48.96 &48.25 &5.26 &4.69 &6.29 &5.41 \\
    AdaptSeg \cite{adaptsegnet}& 47.02 & 69.03 & 66.71 &60.92 &2.95 &3.00 &3.80 &3.25 \\
    SIFA\cite{sifa} & 63.63 & 77.57 & 61.58 &67.59 &2.50 & 3.00 &3.76 &3.09 \\
    Ours & 71.15 & 86.95 & 75.64 & 77.91 &1.11 &1.01 &1.58 &1.23 \\
  \bottomrule
    
  \end{tabular}
  \label{dice and assd}
\end{table*}
\subsection{Performance and Comparison}
We compare our method with some popular unsupervised domain adaptations methods.
\begin{itemize}
    \item [$\bullet$]Unet(supervised) \cite{unet}: We trained a Unet model with labeled targeted data. This result could be \textit{upper bound} for unsupervised domain adaptation. Note that this is a strong baseline for our experiments.
    \item [$\bullet$]Unet(no adaptation) \cite{unet}: We trained a Unet model with labeled source data and directly test on the target data without any domain adaptation.
    \item [$\bullet$]AdaptSeg \cite{adaptsegnet}: This is one of the state of the art methods for unsupervised domain adaptation. AdaptSeg aligned data for source domain and target domain on feature level. For a fair comparison, we also used pre-trained model of Deeplabv2-ResNet101 on ImageNet \cite{imagenet} in experiments, the same as original paper.
    \item [$\bullet$]SIFA \cite{sifa}: SIFA is one of the state of the art unsupervised domain adaptation methods for medical image segmentation. Different from AdaptSeg, SIFA aligned data on images levels and feature levels.
\end{itemize}

We reimplemented all methods with the original public author's code. Note that we referenced our model as SGDR for simplicity, the acronym for Semantic-guided Disentangled representation. We present quantitative results in \cref{dice and assd} and the visual comparisons are shown in \cref{visual comparsion}.

It could be seen that our method has a significant improvement compared to Unet without domain adaptation from \cref{dice and assd}, which has 28\%, 23\%, 22\% higher Dice than comparison methods on three tasks, respectively. 
More importantly, our model also outperforms the current state of the art methods on three tasks, especially SIFA \cite{sifa}. Our model has a 7.22\%, 2.8\%, 3\% higher dice score than SIFA and 1.11, 1.18, 0.19 lower ASSD values than SIFA, respectively. The quantitative results demonstrate the effectiveness of our model.

\cref{visual comparsion} shows the visual results of all comparison methods.
We show the results of three cases of all the compared methods in three tasks. 
It is observed that Unet \cite{unet} trained on source domain without any domain adaptation can hardly predict the cardiac structure correctly.
By using some domain adaptation methods, such as AdaptSeg \cite{adaptsegnet} (4th column) or SIFA \cite{sifa} (5th column) can recover segmentation prediction partlly, but there are still some wrong predictions and unclear boundaries.
Specically, our results are closer to the ground truth and less wrong predictions than other methods, which shows that our model is more robust and accurate.
\subsection{Ablation Study}
Our proposed method uses semantic-guided strategy to improve the performance. In the section, we study the effectiveness of content discriminator, feature discriminator and loss $\mathcal{L}^v_{seg}$ on MR-CT task. The quantitative results are reported in \cref{ablation}, where 'w/o' means to remove the module from our implements. First, the content discriminator encourages encoders to capture domain-invariant features by adversarial training, which improve the performance of domain adaptation by a large margin. Second, we use adversarial training in the semantic space between target images and synthetic target images. We remove the feature discriminator from our model in ablation study. The results show that the feature discriminator can effectively increase average Dice from 72.4\% to 80.18\%. Finally, we evaluated the effectiveness of loss $\mathcal{L}^v_{seg}$.In this work, we use two segmentation loss,$\mathcal{L}^x_{seg}$ and $\mathcal{L}^v_{seg}$. The loss function $\mathcal{L}^v_{seg}$ could further constrain the encoder to exact the anatomical structure of source domain more accurately and get 2.38\% improvement. Overall, the ablation study shows that the semantic-guided strategy of our model achieves better performance and contributes to improvement jointly.
\begin{table}
  \centering
  \caption{Evaluate the effectiveness of three components of our method on MR-CT task.
  'w/o' means remove this component from our implement.
  We report average Dice score of ablation study.}
  \begin{tabular}{lc}
  \toprule
  Method & Average Dice(\%) \\
  \hline
  No adaptation & 52.24 \\
  SGDR w/o content discriminator & 69.6 \\
  SGDR w/o feature discriminator & 72.4 \\
  SGDR w/o loss $\mathcal{L}^v_{seg}$ & 77.8 \\
  SGDR & 80.18 \\
  \bottomrule
  \end{tabular}
\label{ablation}
\end{table}

\section{Conclusion}
We present a novel semantic-guided disentangled representation model for unsupervised domain adaptation of cross-modality medical image segmentation. Our model uses semantic-guided strategy to exact the meaningful domain-invariant features for segmentation task. We apply a content discriminator to help encoders exact domain-invariant features and a feature discriminator to exact the meaningful representation of source domain and target domain. Moreover, an auxiliary segmentation loss is used to keep the anatomical structure and help us exact more meaningful information. We evaluated our method on two public datasets and the experiment results show that our framework improves the performance of segmentation task in target domain by a large margin. Our model can also be readily extended to 3D medical image segmentation task in unsupervised domain adaptation setting, although our work is demonstrated on 2D data.
{\small
\bibliographystyle{ieee_fullname}
\bibliography{egbib}
}

\end{document}